\documentclass[epj-ND]{svjour}
\usepackage[tbtags]{amsmath}
\usepackage{amssymb}
\usepackage[pdftex]{graphicx}

\usepackage{txfonts}
\def\chem#1#2{$\rm{}^{#1}\kern-0.8pt#2$}
\def\reac#1#2#3#4#5#6{$\rm\,{}^{#1}\kern-0.8pt{#2}\,({#3}\,,{#4})\,
{}^{#5}\kern-0.8pt{#6}\,$}
\def\gsimeq{\,\,\raise0.14em\hbox{$>$}\kern-0.76em\lower0.28em\hbox  
{$\sim$}\,\,}  
\def\lsimeq{\,\,\raise0.14em\hbox{$<$}\kern-0.76em\lower0.28em\hbox  
{$\sim$}\,\,}  
\def\beqy{\begin{eqnarray}}
\def\eeqy{\end{eqnarray}}
\def\bmlet{\begin{mathletters}}
\def\emlet{\end{mathletters}}

\begin{document}

\title{From the microcosm of the atomic nuclei to the macrocosm of the stars}

\author{M. Arnould    \inst{} 
\thanks{Presenting author, \email{marnould@astro.ulb.ac.be}}
 \and M. Katsuma   \inst{} 
}

\institute{
     Institut d'Astronomie et d'Astrophysique, Universit\'e Libre de Bruxelles, CP 226, Brussels, Belgium
}

\abstract{A necessary condition for the reliable modelling of the structure or evolution of the stars and of their concomitant nucleosynthesis is the availability of good quality nuclear data in a very wide area of the chart of nuclides. This short review presents a non-exhaustive list of nuclear data of astrophysics interest (masses, $\beta$-decays, thermonuclear and non-thermonuclear reaction rates) for nuclides at the bottom of the valley of nuclear stability (mainly involved in the modelling of non-explosive phases of stellar evolution), or for more or less highly exotic nuclides  (to be considered in the description of stellar explosions). Special emphasis is put on the importance of providing quality nuclear data bases that can be easily used by astrophysicists.}
 
\maketitle

\section{Introduction}
\label{intro}

\begin{figure}[t]
\vskip+0.15cm
\hskip-0.30cm
\resizebox{1.07\columnwidth}{!}{%
\includegraphics{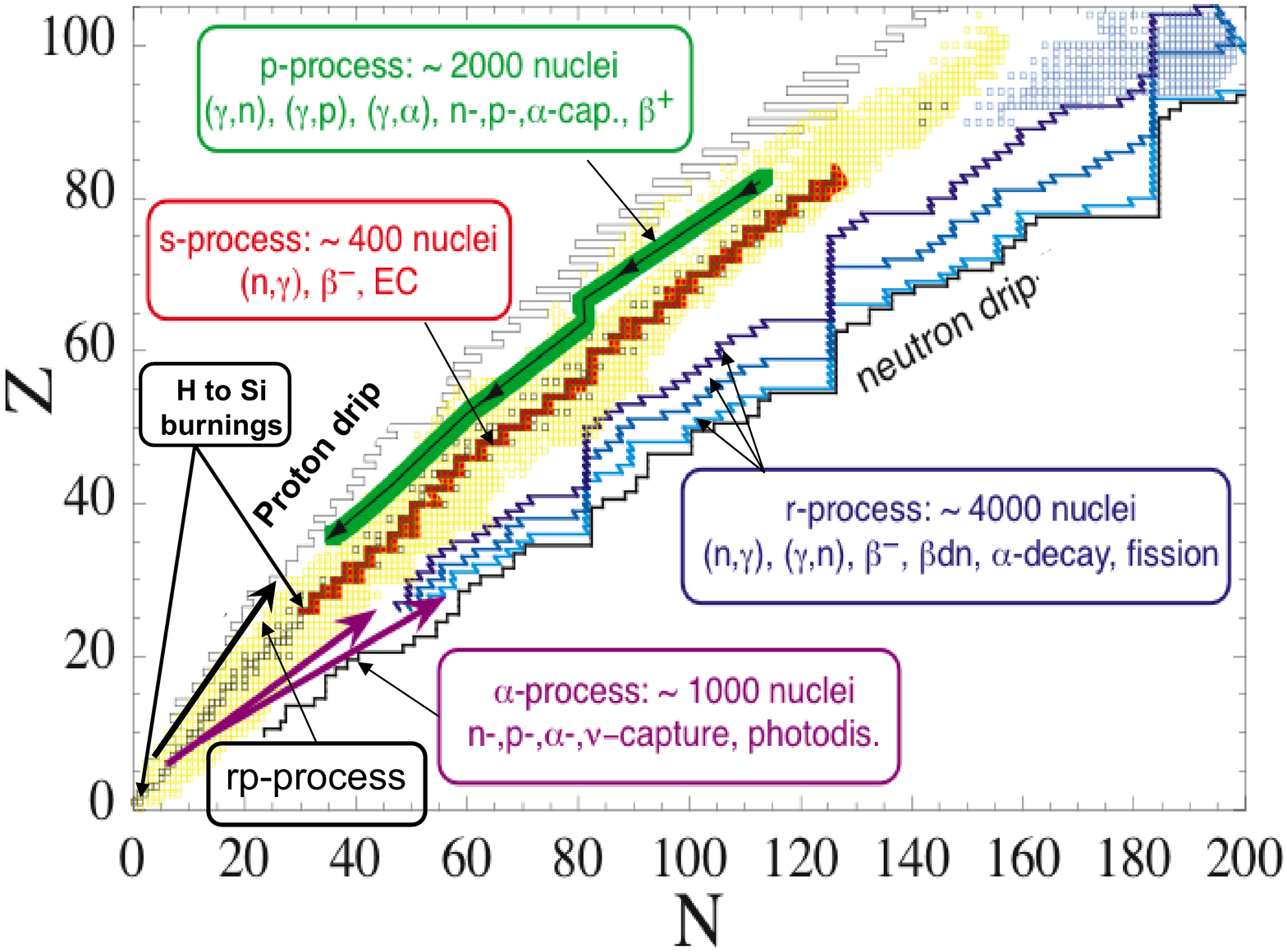}
}
\caption{Astrophysics pushes nuclear physics to the wall. Most of the chart of the nuclides indeed needs to be explored in order to model the displayed nucleosynthesis processes (see the electronic version of the paper for a colour version of this fig.).}
\label{nuclides_chart}
\end{figure}

The close relationship between nuclear physics and astrophysics comes about because the Universe is pervaded by nuclear physics imprints at all scales. Over the years, an impressive body of nuclear data of astrophysics interest have been obtained through laboratory efforts, and many more are expected to accumulate in the future. However, theoretical developments still have -- and will certainly have for a long time to come -- to complement them because very many highly unstable `exotic' nuclei that cannot be produced in the laboratory are expected to be involved in the modelling of a large variety of astrophysics processes and events (see fig.~\ref{nuclides_chart}). Even when laboratory-studied nuclei are considered, theory
has very often to be called for assistance because of the specificity of the physical conditions encountered in stars.
 
The rapidly growing volume  of nuclear data, especially experimental ones, is less and less easily accessible to the astrophysics community. Mastering this volume of information and making it available in an accurate and usable form  is needed.  Compilations and libraries of nuclear data have now been produced for this purpose. 

\section{At the bottom of the valley: Exploration of the world of `almost no event'}
\label{no_event}

The non-explosive nuclear history of a star is made of the burnings of H, He, C, Ne, O, and Si, at least if the stellar mass exceeds about $10 M_\odot$, where $M_\odot$ is the solar mass (see e.g. \cite{arnould99}). These burnings involve essentially stable nuclides only.  Even so, the experimental determination of their cross sections faces enormous  problems \cite{rolfs88}. This situation relates directly to the fact that the energies  of astrophysical interest for charged-particle induced reactions, the so-called Gamow window, are much lower than the Coulomb barrier, with  the consequence that the cross sections can dive into the nano-barn to pico-barn abysses. Thanks to their impressive skill and painstaking efforts, experimental physicists involved in nuclear astrophysics have been able to provide the smallest nuclear reaction cross sections ever measured in the laboratory by pushing technologies to their limits. In spite of this,  with a few exceptions only, they have not succeeded yet in reaching the region of `almost no events' of astrophysical relevance. Indirect approaches are a very important complement to direct measurements. From experiments on reactions of no astrophysical relevance, information (e.g. resonance characteristics) can be gained on reactions of interest (\cite{arnould99} for a brief overview). Theorists are thus requested to supply reliable extrapolations from the experimental data obtained at the lowest possible energies, or to evaluate cross sections from indirectly gained information. Electron screening effects in laboratory conditions, or the contribution to the reaction rates of the target excited states, bring their share of additional difficulties (e.g. \cite{arnould99}).

\subsection{Compilation of thermonuclear reaction rates}
\label{comp_reaction}

As said above, the establishment of the required level in the communication between nuclear physicists and astrophysicists makes necessary the build-up of well-documented and evaluated sets of experimental data of astrophysical relevance. This philosophy has been the driving motivation for the construction of the {\it Nuclear Astrophysics Compilation of REaction rates for astrophysics} (NACRE) \cite{angulo99} aimed at superseding the work of Fowler and collaborators  \cite{caughlan88}.  Slightly more than half of the thermonuclear reaction rates on mostly stable targets up to Si considered in \cite{caughlan88} have been re-compiled on the basis of a careful evaluation  of experimental data available up to 15 June 1998. NACRE is accessible electronically through the website {\it http://www.astro.ulb.ac.be}, where material not published by \cite{angulo99} is available.   In particular, reaction rates are given in tabular form for more extended temperature grids.  

Other astrophysics-oriented experimentally-based thermonuclear reaction rate compilations have appeared after the NACRE publication.  References and additional data or evaluations of relevance published up to the beginning of 2005 can be found in \cite{aikawa05}. These data and still more recent ones will be evaluated and integrated into an updated and extended new version of NACRE, referred to as NACRE II (see sect.~\ref{NACREII}).

\begin{table}[t]
\centering
\caption{Oxygen isotopic composition at the surface of stars of different masses $M$ and of initial solar content after transport of part of the central H-burning ashes to the surface by the stellar envelope convection. The isotopic ratios and the quoted uncertainties result from the use of the recommended NACRE rates of H burning through the CNO mode, and of their lower and upper limits provided by NACRE (from \cite{herwig03}).}
\label{tab:1}       
\begin{tabular}{lll}
   \hline\noalign{\smallskip}
   $M (M_\odot)$ & \chem{16}{O}/\chem{17}{O} & \chem{16}{O}/\chem{18}{O} \\
   \noalign{\smallskip}\hline\noalign{\smallskip}  
  1.0 & 2410$^{+13}_{-16}$ & 469$^{+46}_{-12}$ \\
  1.5 & 1260$^{+180}_{-130}$ & 545$^{+70}_{-20}$ \\
  2.0 & 115$^{+38}_{-20}$ & 565$^{+95}_{-20}$ \\
  Initial & 2465 & 442 \\
   \noalign{\smallskip}\hline
\end{tabular}
\end{table}

\begin{figure}
\vskip-0.9cm
\hskip-0.5cm
\resizebox{1.1\columnwidth}{!}{%
 \includegraphics{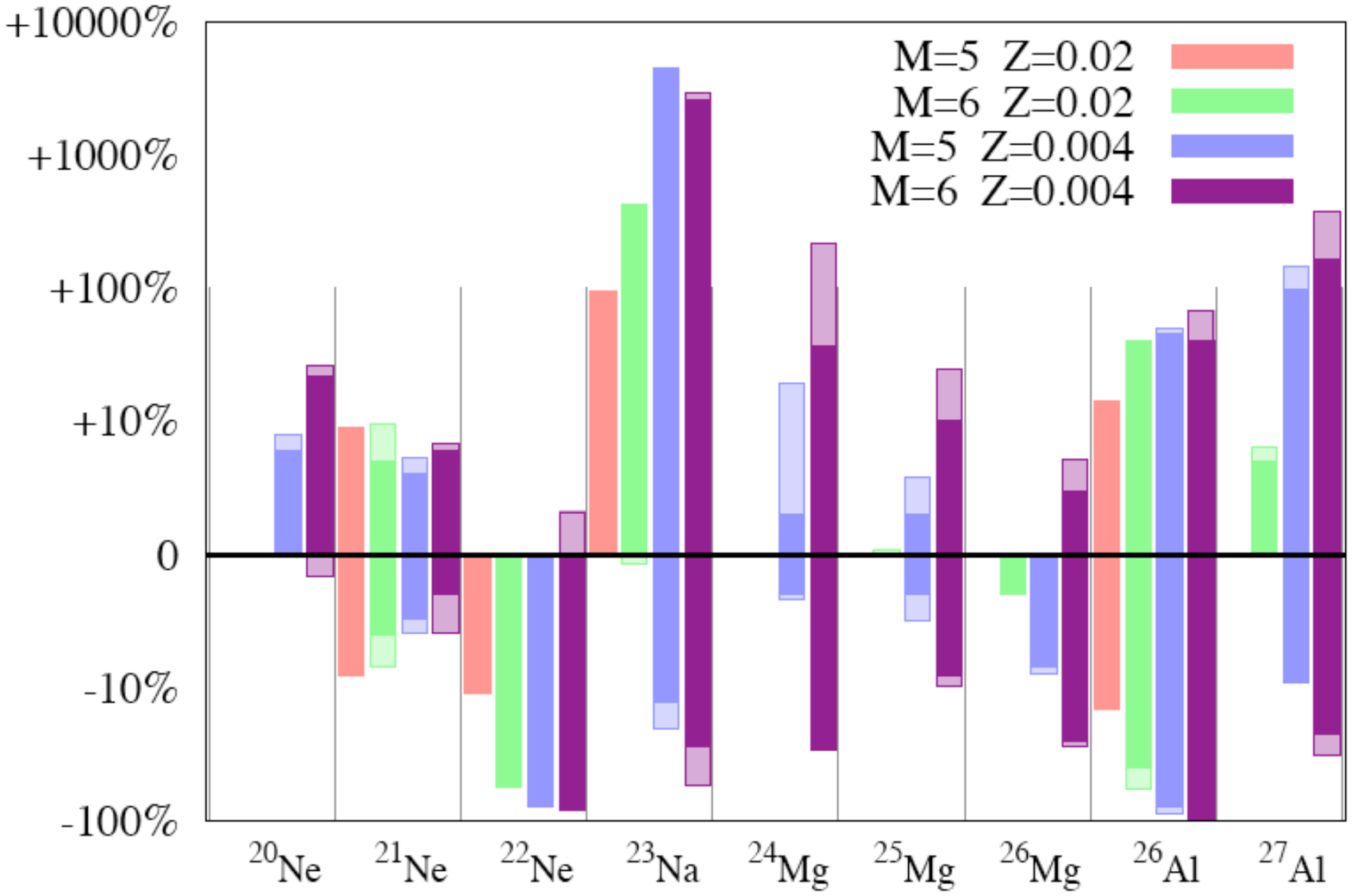} 
}
\caption{Deviations of the yields of the nuclides involved in the NeNa and MgAl chains from reference values obtained with recommended reaction rates when these rates are varied within evaluated limits. The different grey scales correspond to different stellar masses and/or metallicities (see the electronic version of the paper for a colour version of this fig.) (from \cite{izzard07}).}
\label{yields_ne}
\end{figure}

Various analyses of the astrophysics impact of the uncertainties in compiled reaction rates have been conducted (e.g. \cite{arnould99a,herwig03,izzard07}). Table~\ref{tab:1} and fig.~\ref{yields_ne} illustrate uncertainties of purely nuclear origins associated to the operation of the CNO and NeNa-MgAl modes of hydrogen burning in various stars at different stages of their evolution.

\subsection{NACRE II: an update and extension of the NACRE compilation}
\label{NACREII}

\begin{figure}[b]
  \centering{\includegraphics[width=0.90\linewidth]{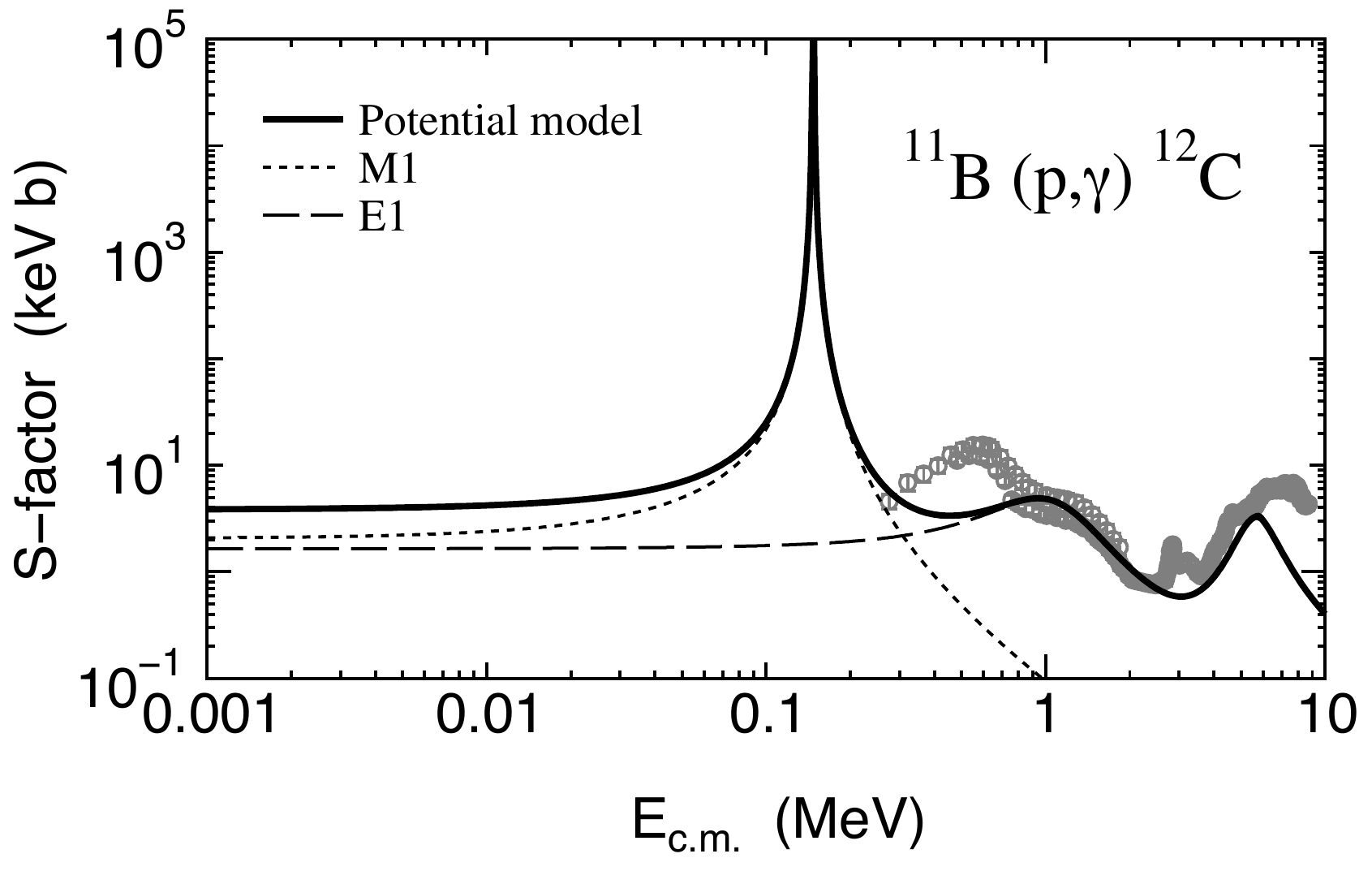}}
  \caption{The $S$-factor for $^{11}$B(p,$\gamma$)$^{12}$C calculated with our potential model (solid curve) compared with the experimental data reported by NACRE~\cite{angulo99} (open dots and thick grey line). The calculated contributions from the E1 and M1 transitions are indicated by the dashed and dotted lines.} 
  \label{fig-b11pg}
\end{figure}

\begin{figure}
  \centering{\includegraphics[width=0.88\linewidth]{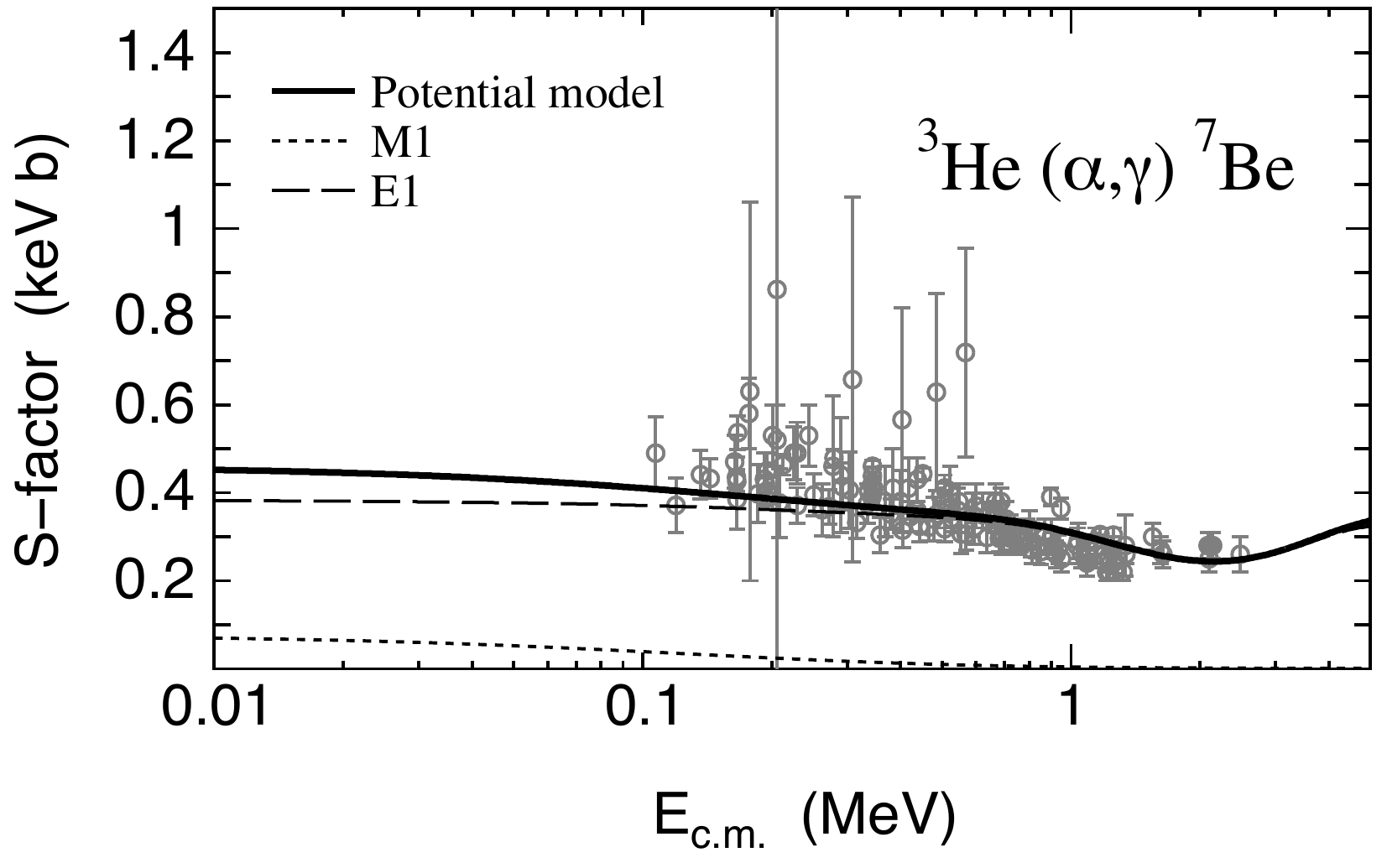}}
  \caption{Same as fig.~\ref{fig-b11pg}, but for $^{3}$He($\alpha$,$\gamma$)$^{7}$Be.}
  \label{fig-he3ag}
\end{figure}

 NACRE II is meant to update NACRE through the inclusion of new experimental data, and to extend it with the evaluation and compilation of data on targets with mass numbers $A > 28$. A NACRE II status report can be found in \cite{katsuma07}, and we limit ourselves here to a very brief account. 

One of the most important features of NACRE II is the development of direct reaction models, like the potential model and distorted-wave Born approximation, for use in the extrapolation of experimental data to the very low energies of astrophysical relevance, special attention being paid to the reliability of these extrapolations.  
 
Figures~\ref{fig-b11pg} and \ref{fig-he3ag} illustrate our analysis of \reac{11}{B}{p}{\gamma}{12}{C} and \reac{3}{He}{\alpha}{\gamma}{7}{Be} in terms of the astrophysical factor $S(E) = E \exp(2\pi\eta)\,\sigma(E)$, where $\eta$ is the Sommerfeld parameter, and $E$ is the centre-of-mass energy in the entrance channel.

Figure~\ref{fig-b11pg} shows that  the trend of the experimental data for $E\ge$ 1 MeV is reasonably well reproduced with the potential model. The resonances are obtained without the inclusion of additional Breit-Wigner contributions. At lower energies, a $J^\pi=2^+$ resonance at $E=$ 0.149 MeV (\chem{12}{C} excitation energy $E_{\rm x}=$ 16.105 MeV) has been reported, though the cross section has not been measured. The potential strength of our model is chosen in order to reproduce this resonance energy. The extrapolation down to low energies includes the transitions to the \chem{12}{C} ground (0$^+$) and 1st excited (2$^+$, $E_{\rm x}=$ 4.438 MeV) states, but neglects other states below the proton threshold. In addition, only the E1 and M1 contributions are taken into account. From the preliminary results displayed in fig.~\ref{fig-b11pg}, the E1 and M1 contributions are seen to contribute significantly to the low-energy $S$-factor. This expectation could be amenable to an experimental check through a measure of  the level of anisotropy of the produced $\gamma$-rays. The  \reac{11}{B}{p}{\gamma}{12}{C} rates deduced from the displayed $S$-factor  are expected to exceed those reported in NACRE for temperatures $T \ge 10^8$ K .
 
Figure~\ref{fig-he3ag} shows that the non-resonant $S$-factor for $^{3}$He($\alpha$,$\gamma$)$^{7}$Be can be nicely reproduced by our potential model, at least in the limits of sometimes large experimental uncertainties. The relative orbital angular momentum of $^3$He and the $\alpha$-particle is assumed to be the p-wave for the final bound states, so that the s-wave and p-wave of the incident $\alpha$-particle contribute to the E1 and M1 transitions. The transitions to the ground (3/2$^-$) and 1st excited (1/2$^-$, $E_{\rm x}=$ 0.429 MeV) states in $^{7}$Be are taken into account. From the derived low-energy $S$-factor, it is concluded that the resulting \reac{3}{He}{\alpha}{\gamma}{7}{Be} rate does not deviate significantly from the NACRE one.

Figures~\ref{fig-b11pg} and \ref{fig-he3ag} illustrate that our new extrapolation procedure applied to the same experimental data as in NACRE can produce reaction rates that are similar or more or less different from NACRE. Additional deviations are obtained when new experimental data are taken into account. A more complete account of the developed models and of the update of the compiled NACRE II rates will be published elsewhere.
  
\section{At the bottom of the valley: Boating on the s-process neutron flow}
\label{s-process}
 
The s(`slow')-neutron-capture process relies on the assumption that pre-existing (`seed') nuclei are exposed to a flux of Maxwellian-distributed neutrons that is weak enough for allowing a $\beta$-unstable nucleus produced by a (n,$\gamma$) reaction to decay before being able to capture a neutron. As a result, the associated nuclear flow (the `s-process path') is constrained to the close vicinity of the line of $\beta$-stability (see fig.~\ref{nuclides_chart}) with an endpoint located at \chem{209}{Bi}.  
 
The main reactions responsible for the production of the necessary neutrons are \reac{13}{C}{\alpha}{n}{16}{O} and \reac{22}{Ne}{\alpha}{n}{25}{Mg}, the relative importance of which depends on the stellar mass and initial composition. Both astrophysical and nuclear physics uncertainties still affect the predicted neutron production efficiency. A discussion of the still controversial situation concerning the \chem{22}{Ne} $\alpha$-particle capture rates can be found in \cite{karakas06}.

Most thermonuclear (n,$\gamma$) rates needed for the s-process modelling have been measured with often great accuracy, at least at $kT \approx 30$ keV \cite{bao00}. Some further measurements would be welcome in certain mass ranges ($12 \lsimeq A \lsimeq 100$; A $>$ 190) or down to energies around $kT = 5$ keV. The rates of $\beta$-decay of nuclei just neighbouring the bottom of the valley of stability are the necessary nuclear complement in the study of the s-process. Their evaluation raises some problems  \cite{takahashi87}.

\section{Climbing the slopes of the valley of stability in search of the nuclear exoticism}
\label{exotism}

As illustrated in fig.~\ref{nuclides_chart}, the `rapid proton capture' (rp)-process and the so-called p-process involve more or less exotic neutron-deficient nuclides, while the so-called $\alpha$-process and the `rapid neutron capture' (r)-process develop on the neutron-rich side of the valley of stability. In addition, the H- to Si-burning stages that may accompany non-explosive stellar evolutionary phases may also develop during higher-temperature exploding stages, and may slightly overflow the banks of the valley of nuclear stability.

The modelling of these various processes requires a very large body of properties to be known for a huge variety of light to heavy nuclei from close to the proton-drip line to close to the neutron-drip line. No doubt that this represents a really tough challenge for experimental and theoretical nuclear physics. This challenge concerns ground and excited state properties (masses, shapes, pairing energies, spectra of excited states and nuclear level densities, spontaneous $\alpha$- and $\beta$-decay or fission probabilities, ...), as well as the properties of their interactions mainly with nucleons  or $\alpha$-particles, as well as with photons.  
A great experimental effort is made in many places to take up as much of the challenge as possible.  On the theoretical side, one of the basic aims is to evolve from phenomenological and more or less highly parametrized approaches to microscopic descriptions, the ultimate goal being the coherent microscopic treatment of all the properties of all the nuclei through the use of a  microscopic model that is universal and global to the largest possible extent. For a discussion of the importance of developing such models, the reader is referred to e.g. \cite{arnould06}.

\subsection{Nuclear masses and other structure properties}
\label{masses}

Recently, impressive progress has been made in the measurement of the masses of unstable nuclei (e.g. \cite{lunney03}), and compilations are regularly published \cite{audi03}, with additional news to be found in  the `Atomic Mass Data Center' Bulletin. Much more is expected to come in the near future. This effort goes along with important theoretical advances culminating with the construction of complete mass tables based on the microscopic Hartree-Fock-Bogoliubov (HFB) model (Goriely et al., these proceedings). The good news are that the resulting predictions can reach the accuracy of the microscopic-macroscopic models, while their reliability far from stability is expected to be better. The HFB model can also predict self-consistently a variety of other structure properties (like radii) with a very good accuracy.

\begin{figure}[t]
\vskip-0.1cm
\hskip-0.08cm
\resizebox{0.95\columnwidth}{!}{%
 \includegraphics{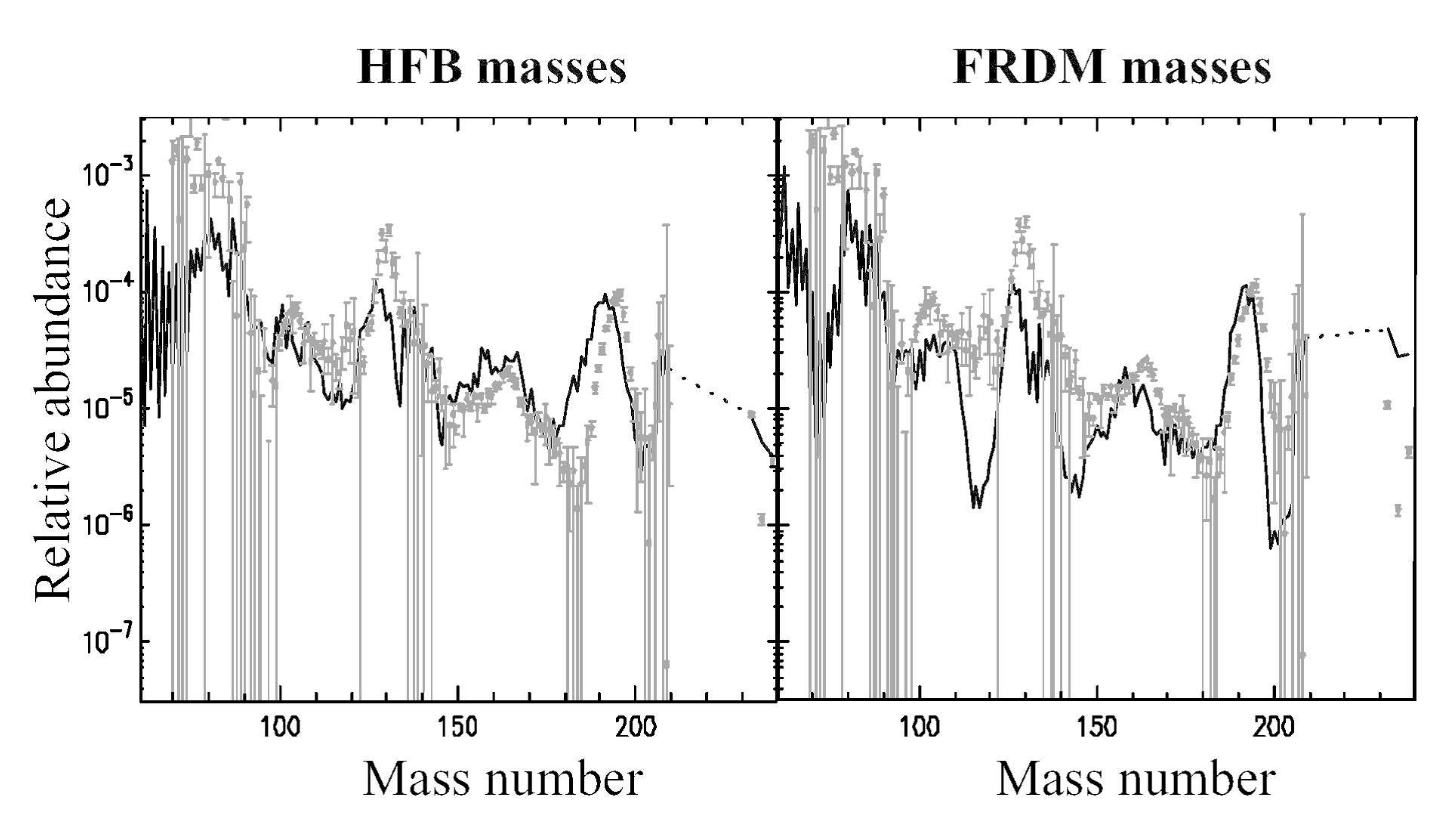}  
}
\vskip-0.1cm
\caption{Impact of different mass predictions (left panel: microscopic HFB model; right panel: macroscopic-microscopic FRDM model) on the yields from a prompt explosion r-process. The solar r-abundance are displayed for comparison (light-grey dotted line with vertical bars representing uncertainties; see \cite{arnould07} for details). The predicted abundances and the solar data are normalised at the solar r-process peak located at \chem{195}{Pt} (from \cite{wanajo05}).}
\label{yields_mass_prompt}
\end{figure}

\begin{figure}
\vskip-0.2cm
\hskip+0.0cm
\resizebox{0.90\columnwidth}{!}{%
 \includegraphics{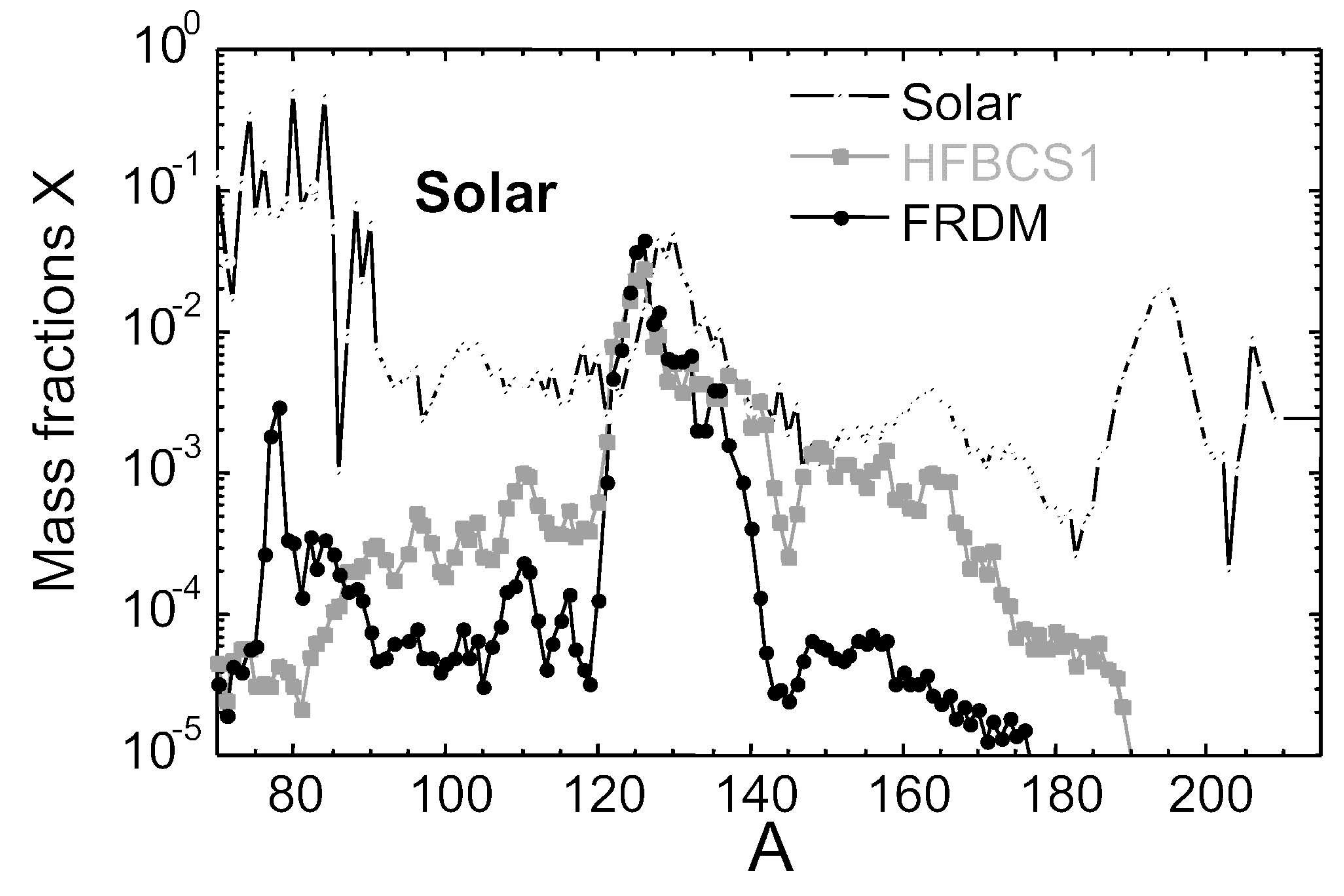}  
}
\vskip-0.2cm
\caption{Impact of two different mass predictions (grey squares: HFBCS; black dots: FRDM) on the r-nuclide yields from a `delayed' supernova explosion. The thin dashed line displays `standard' solar r-nuclide abundances \cite{arnould07} (uncertainties not shown). The calculated abundances are normalised to the solar $A = 130$ r-peak (by courtesy of S. Goriely). }
\label{yields_mass_wind}
\end{figure}

\begin{figure}[t]
\vskip+0.1cm
\hskip-0.225cm
\resizebox{0.90\columnwidth}{!}{%
 \includegraphics{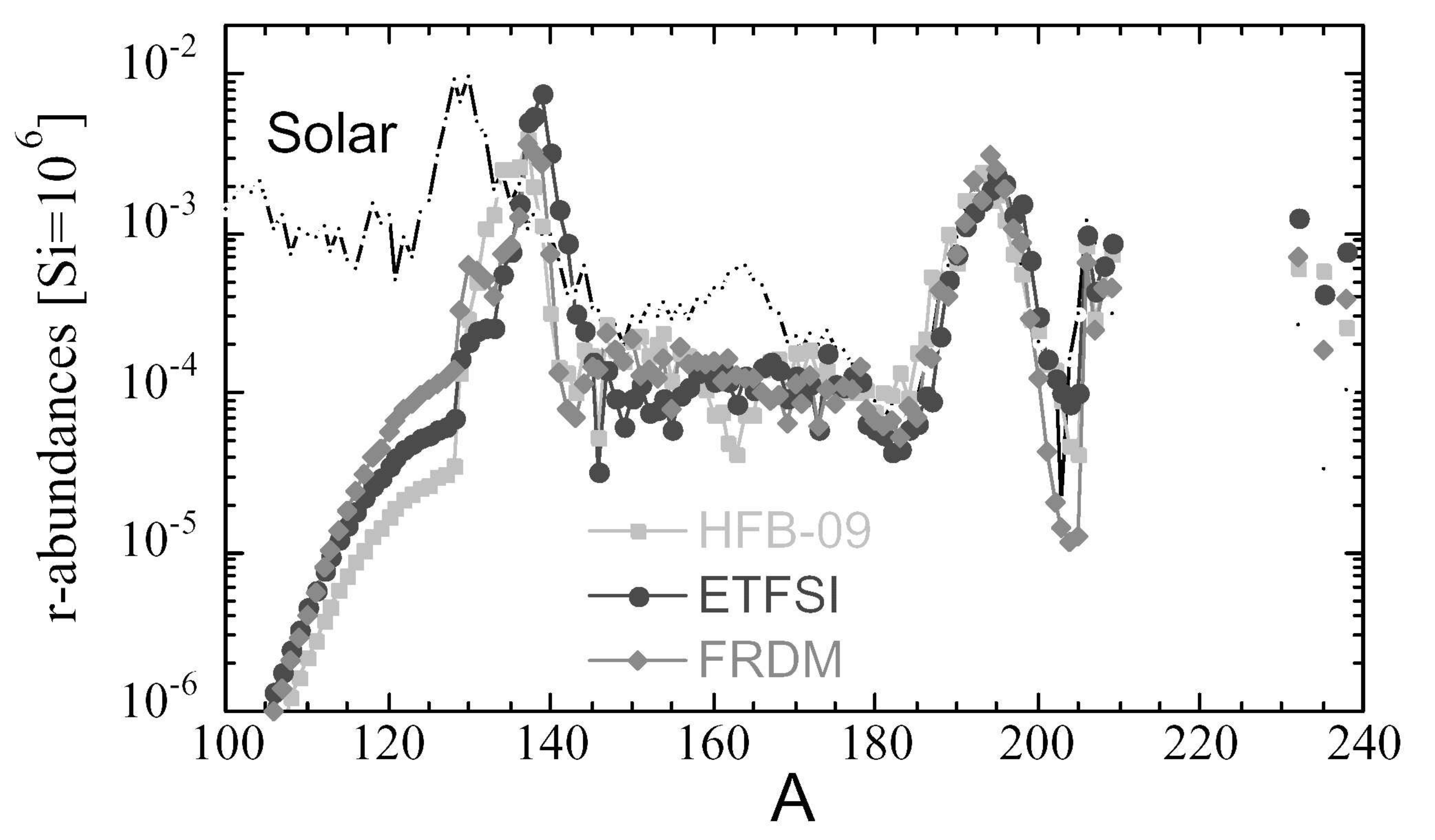}  
}
\vskip-0.2cm
\caption{Impact of different mass predictions (light-grey squares: a HFB model; black dots: ETFSI; grey diamonds: FRDM) on the r-nuclide yields from the decompression of cold neutron matter following the coalescence of two neutron stars. The thin dotted line displays `standard' solar r-nuclide abundances \cite{arnould07} (uncertainties not shown). The calculated abundances are normalised to the solar $A = 195$ r-peak (by courtesy of S. Goriely).}
\label{yields_mass_decompression}
\end{figure}

Experimentally unknown masses come into play mainly in the rp-process \cite{schatz06}, and in the r-process. The impact of the mass predictions on the yields of the r-process assumed to develop in various astrophysical scenarios has been analysed in so many places that it is not possible to review these studies here.  Figures~\ref{yields_mass_prompt}, \ref{yields_mass_wind} and \ref{yields_mass_decompression} compare the r-nuclide abundances predicted with different mass models in three astrophysical situations (`prompt' supernova explosion, `delayed' supernova explosion, and decompression of cold neutron star matter resulting from the coalescence of two neutron stars). It is seen that the abundance changes due to the use of different predicted masses has an impact whose extent varies with the astrophysical scenario, being the least pronounced for the decompression of neutron star matter (fig.~\ref{yields_mass_decompression}). From this sensitivity, and from a comparison between solar and predicted r-abundances, some are tempted to evaluate the relative virtues of different mass models. This is quite risky, especially as the proper site(s) for the r-process remain(s) unknown, as recently reviewed by \cite{arnould07}. In fact, figs.~\ref{yields_mass_prompt} - \ref{yields_mass_decompression} are based on models that have been parametrised in one way or another to obtain a well-developed r-process. In reality, no prompt explosion is obtained in detailed simulations, while delayed explosions are concluded to be unsuccessful, even if some hope remains to get supernovae of this type, and possibly some concomitant r-process. In view of these problems, the coalescence of neutron stars appears today as the most plausible site for the r-process, but it also faces some serious astrophysics difficulties \cite{arnould07}. It is remarkable that it is just this scenario that minimises the impact of different mass predictions!

\subsection{Beta-decays}
\label{beta}

Many new $\beta$-decay half-lives have been measured in recent years (see the compilations by \cite{audi03}). Different approaches have been proposed to understand, and wishfully predict, $\beta$-decay rates. The models range from macroscopic approximations, referred to as the `Gross Theory' from which an extended table of $\beta$-decay half-lives has been produced \cite{takahashi73}, to fully microscopic large-scale shell models. Those in between are global approaches of various kinds with a more or less pronounced microscopic character (e.g. \cite{arnould07} for a brief review).

Quite large uncertainties generally affect the $\beta$-decay half-life predictions. As discussed by \cite{arnould07}, the Gross Theory with global parameter values performs remarkably well in its accuracy to reproduce experimental data. One may of course wonder about the reliability of the predictions of this macroscopic approximation very far from the valley of stability. The present level of accuracy of the existing microscopic models is far from being satisfactory, so that their reliability is difficult to assess. In addition, it is still a very long way for microscopic models to produce the large-scale $\beta$-decay data required by astrophysics. 

 Uncertain $\beta$-decays play a role in the rp-process \cite{schatz06}. The changes in the yields of the r-process from the use of different $\beta$-decay models have also been calculated. In view of the present failure of the prompt or delayed supernova scenarios (sect.~\ref{masses}), we do not review these sensitivity analyses here. In the case of the coalescence of neutron stars, the impact of $\beta$-decay uncertainties is found rather limited, as already reported in sect.~\ref{masses} for the masses (S. Goriely, private communication). 
 
\subsection{Thermonuclear nucleon and $\alpha$-particle capture rates, and photo-nuclear reactions}
\label{rates}

The cross sections of a very limited number of reactions involving light enough (mostly $A \lsimeq 26$) neutron-deficient nuclei close enough to stability have been measured directly thanks to the development of Radioactive Ion Beam facilities (e.g. \cite{blackmon05} for a review). In most cases, however, they are evaluated indirectly through the use of different techniques (e.g. \cite{arnould99,champagne92,spitaleri99}).

Far enough from stability and for light enough systems, the lack of useful experimental data combined with the absence of reliable models make cross section predictions especially risky. The theoretical situation improves for intermediate-mass and heavy systems, at least not too close to the drip lines. In such cases, there is hope that a statistical model of the Hauser-Feshbach type might apply, with a rather systematic behaviour of its basic ingredients. The modelling of these ingredients (masses, nuclear level densities, $\gamma$-ray strength functions, optical potentials) has been reviewed briefly by \cite{arnould07}, emphasis being put on progress towards predictions relying on global and universal microscopic models. The latest version of the code TALYS includes many of these advances, as well as others (Koning et al., these proceedings). More has to come, as exemplified by a recent analysis by \cite{goriely07} of the isovector part of the imaginary optical potential which is demonstrated to affect drastically the Hauser-Feshbach predictions of the (n,$\gamma$) rates for very neutron-rich nuclei. 

In addition, non-statistical effects may play an important role in certain situations. When approaching the neutron drip line, this concerns in particular a pre-equilibrium phase in the reaction process (Goriely et al., these proceedings),  or the `direct capture' (DC) mechanism, in which there is little rearrangement of the nucleons among the available states. This situation shows similarities with the one encountered in light stable nuclei, where the formation of an equilibrium compound nucleus is likely to be prevented. In such conditions, radiative captures by nuclei close to the drip lines may be  dominated by direct electromagnetic transitions to bound final states. A simplified DC model for use in astrophysical applications is reviewed by \cite{arnould07}. Its predictions may deviate drastically from the Hauser-Fesbach ones for neutron captures by nuclei with very loosely bound neutrons  \cite{goriely98}. 

Photo-nuclear reactions of the ($\gamma$,n), ($\gamma$,p) or ($\gamma,\alpha$) types induced by black-body photons in stellar interiors also play a very important role in the p-process \cite{arnould03}, or in high-temperature scenarios of the r-process \cite{arnould07}. In recent years, great progress has been made in the direct measurements of some of these reactions, and much more is expected in the future \cite{utsunomiya06}. These measurements provide a very good test of the calculation of the photo-disintegration rates which relies usually on the reciprocity theorem. Its adoption may be invalidated by some effects far from stability (Goriely et al., these proceedings).

Rate uncertainties affect the rp-process \cite{schatz06}, as well as the p-process (see the detailed review by \cite{arnould03}). Similar sensitivity analyses have been carried out for the r-process. The decompression of cold neutron matter shows a limited sensitivity of the ensuing r-process to the rate uncertainties (S. Goriely, private communication). The situation in this respect  with the other scenarios mentioned in sect.~\ref{masses} is not reviewed here, as they are still affected by huge astrophysics uncertainties.

\subsection{Other nuclear processes of astrophysical interest: fission, delayed processes, neutrino-matter interaction}
\label{varia}

There are other processes of astrophysical interest involving exotic nuclei that cannot be reviewed in detail here (see e.g. \cite{arnould07} for more information). One of these is fission, the probability of which remains very uncertain in general, even if some progress has been made on the way to microscopic predictions. Beta-delayed processes (neutron emission or fission) also raise many challenging problems. In addition, the interaction of neutrinos with matter may lead to a variety of important astrophysical processes. Their cross sections depend on the still poorly known global behaviour of the weak interaction strength functions. Let us also note that there have been recent speculations on the role of fissions possibly induced by high-energy neutrino captures. These fission probabilities cannot be calculated at present with any degree of reliability.

\subsection{BRUSLIB: The Brussels nuclear physics library for astrophysics}
\label{bruslib}

Mastering the huge volume of nuclear information described above in this section, and making it available in an accurate and usable form for incorporation into astrophysics models is clearly of pivotal importance. The recognition of this necessity  has been the driving motivation for the construction of the BRUSsels LIBrary (BRUSLIB) of computed data of astrophysics relevance. It provides  in an easily accessible electronic form at the address {\it http://www.astro.ulb.ac.be}  extended tables of masses, nuclear level densities and partition functions, fission barriers, and about 100000 thermonuclear stellar rates calculated within the Hauser-Feshbach model for nucleon- and $\alpha$-particle captures, as well as for photo-induced reactions. In addition to the unprecedented broadness of its scope, BRUSLIB has the unique and most important feature of relying to the largest possible extent on global and coherent microscopic nuclear models.  For completeness, BRUSLIB also contains the experimentally-based NACRE rates (Sect.~\ref{comp_reaction}). More details about BRUSLIB can be found in \cite{arnould06}.

BRUSLIB is complemented with a package called NETGEN for the construction of nuclear reaction networks based on the nuclear physics input from BRUSLIB and, when necessary, from other sources.  A more complete description of NETGEN can be found in \cite{aikawa05}.

\section{Non-thermonuclear reactions}
\label{spallation}

Non-thermonuclear reactions at ultra-high or high energies (in excess of about 1 GeV up to about $10^{12}$ GeV/nucleon), as well as at medium energies (from well above the Coulomb barrier to about 1 GeV/nucleon)\footnote{These energy ranges are just indicative, and cannot be defined unambiguously for all reactions} play a very important role in many astrophysical situations, as in a large variety of other scientific applications.  

The interaction of ultra-high-energy cosmic rays of likely extra-galactic origin with the cosmic microwave background radiation (CMB) involves not only pion photo-production from protons, but possibly also the photo-disintegration of a large variety of heavy nuclides.  Only a limited set of experimental cross sections for the relevant photo-disintegrations and the energy range of interest for this problem (typically up to 50 MeV or so) are known experimentally \cite{iaea00}, so that most of them have to be evaluated (see \cite{khan05}). The code TALYS \cite{talys} has been used by \cite{khan05} to calculate a large variety of photo-reaction channels of interest  (emission of one or multiple nucleons and/or $\alpha$-particles) for all nuclides with $12 \le A \le 56$.

At high energies, spallation reactions are essential in the shaping of the elemental and isotopic composition of the relativistic galactic cosmic rays (GCRs), and play a central role in `propagation models' attempting to determine the GCR abundance variations as the result of the interaction of about half of the GCR $Z \ge 6$ nuclides  with the interstellar matter when they travel from the GCR source(s) to the Earth (e.g. \cite{vernois96}). An important subset of the predictions from the propagation models concerns the production by spallation of the galactic Li, Be and B content (e.g. \cite{reeves94}). Experimental data (see \cite{binns05,silberberg98} for references) have accumulated over the years, but still need to be complemented with theoretical evaluations. Cosmic-ray physicists routinely make use of parametric semi-empirical cross sections \cite{silberberg98}. It would clearly be of interest to develop reliable and more physical models than these for the evaluation of spallation cross sections of astrophysics interest.
 
Various non-exploding  or exploding stars are  known to accelerate particles at medium energies. These `stellar energetic particles' can interact with the material (gas or grains) at the stellar surfaces, in circumstellar shells, or in the local ISM. The Sun itself produces such particles, mainly in the 10 to 200 MeV/nucleon range.

 At medium energies, pre-equilibrium, spallation and fragmentation reactions occur, and can produce a large variety of stable and radioactive residuals. This production by solar energetic particles is of importance not only for astrophysics, but also for a large variety of applications (e.g. space and environmental sciences, aviation technology) \cite{michel07}. The composition of the envelope of certain chemically peculiar magnetic stars, and in particular the suggested presence of radioactive elements (like Tc, Pm or certain $84 \le A \le 99$ actinides), could also find an explanation in the interaction of medium-energy particles with the surface material \cite{goriely07a}.  Much progress in our knowledge of the cross sections of interest has been made in recent years. The dedicated experimental effort is complemented with the development of models and computer codes like TALYS \cite{talys,michel07} (see also Koning et al., these proceedings). However, open problems remain, like the theoretical description of the production of intermediate-mass fragments, or the yields of neutron-induced reactions above about 30 MeV, for which measurements are scarce and models not yet reliable enough. 

\begin{acknowledgement}
We thank K. Takahashi for his contribution to NACRE II. This work has been supported in part by the Belgian Inter-university Attraction Pole PAI 5/07, and by a Konan University-Universit\'e Libre de Bruxelles Convention.  
 \end{acknowledgement}

\end{document}